\documentclass{article}

    \PassOptionsToPackage{numbers, compress}{natbib}
    \usepackage[preprint]{neurips_2025}

\usepackage[utf8]{inputenc} % allow utf-8 input
\usepackage[T1]{fontenc}    % use 8-bit T1 fonts
\usepackage{hyperref}       % hyperlinks
\usepackage{url}            % simple URL typesetting
\usepackage{booktabs}       % professional-quality tables
\usepackage{amsfonts}       % blackboard math symbols
\usepackage{nicefrac}       % compact symbols for 1/2, etc.
\usepackage{microtype}      % microtypography
\usepackage{xcolor}         % colors
\usepackage[pdftex]{graphicx} 
\usepackage{multirow}
\usepackage{booktabs}

\title{Rethink Domain Generalization in Heterogeneous Sequence MRI Segmentation}

\author{%
 Zheyuan Zhang$^{1}$\qquad Linkai Peng$^{1}$ \qquad Wanying Dou$^1$ \qquad Cuiling Sun$^1$ \And Halil Ertugrul Aktas$^1$ \qquad Andrea M. Bejar$^1$ \qquad Elif Keles$^1$ \And Gorkem Durak$^1$ \qquad Ulas Bagci$^1$ \\
    $^1$ Department of Radiology, Northwestern University, Chicago, IL, USA
}

\begin{document}

\maketitle

\begin{abstract}

Clinical magnetic‑resonance (MR) protocols generate dozens of T1 and T2 sequences whose visual appearance differs more than the acquisition sites that produce them.  Existing domain‑generalization benchmarks focus almost exclusively on \textit{cross‑center} shifts and overlook this dominant source of variability. Pancreas segmentation remains a major challenge in abdominal imaging: the gland is small, irregularly shaped, surrounded by stomach, duodenum and visceral fat, and often suffers from low T1 contrast. State‑of‑the‑art deep networks that already achieve >90\% Dice on the liver or kidneys still miss 20–30\% tissue on the pancreas. In addition, the organ is systematically under‑represented in public cross‑domain benchmarks—even though accurate pancreas delineation is critical for early cancer detection, surgical planning and diabetes research. To close this gap, we present \textit{PancreasDG}, a large-scale multi-center 3D MRI pancreas segmentation dataset specifically designed to investigate domain generalization in medical imaging. The dataset comprises 563 MRI scans from six institutions, spanning both venous phase and out-of-phase sequences, enabling systematic study of both cross-center and cross-sequence variations with pixel‑accurate pancreas masks created by a double‑blind, two‑pass protocol. Through comprehensive analysis of this dataset, we reveal three critical insights: (i) limited sampling introduces significant variance that may be mistaken for distribution shifts, (ii) cross-center performance correlates with source domain performance for identical sequences, and (iii) cross-sequence shifts present fundamentally different generalization challenges requiring specialized solutions. We further propose a semi-supervised learning approach for cross-sequence generalization that leverages anatomical invariances, significantly outperforming state-of-the-art domain generalization techniques with +61.63\% Dice score improvements  (from 43.55\% to 70.39\%) and +87.00\%  (from 35.62\% to 66.61\%) on two independent test centers for cross-sequence segmentation. \textit{PancreasDG} establishes a new benchmark for domain generalization in medical imaging with implications beyond medical applications. Dataset, code, and models will be available at \url{https://pancreasdg.netlify.app}..

\end{abstract}

\section{Introduction}
\label{sec:intro}

\begin{figure}[htbp]
    \centering
    \includegraphics[width=0.6\linewidth]{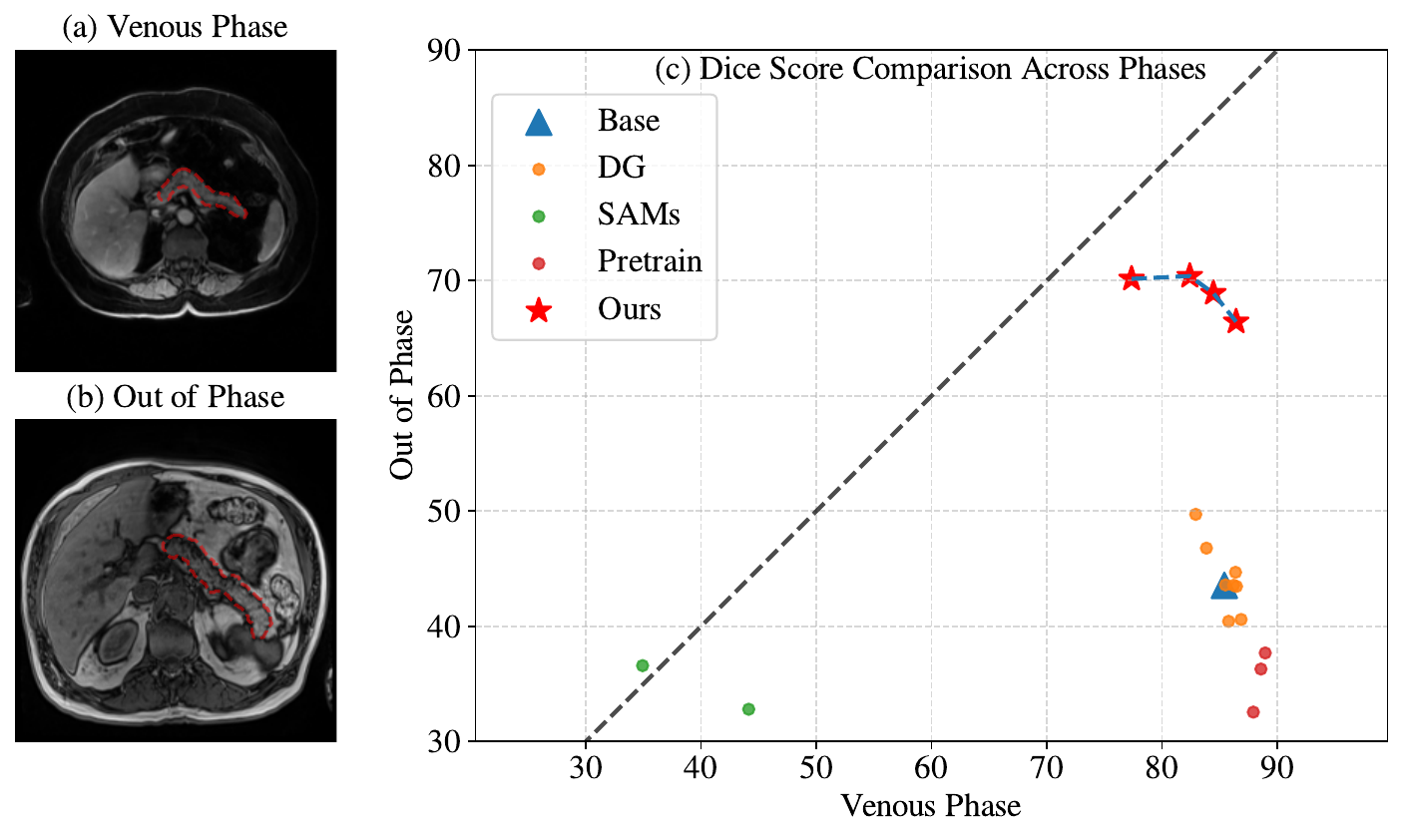}
    \caption{\textit{PancreasDG} reveals the substantial impact of sequence variations on segmentation performance between (a) venous and (b) out-of-phase MRI. Unlike center shifts, sequence variations present a more formidable challenge due to the vast diversity of clinical protocols. (c) Our semi-supervised pretraining approach significantly outperforms existing domain generalization methods and large segmentation models when trained only on venous phase data.}
    \label{fig:domain_comparision}
    \vspace{-4mm}
\end{figure}

Medical image segmentation plays a crucial role in modern clinical applications, including organ delineation, disease detection, and treatment planning \cite{ronneberger2015unet,isensee2021nnunet,hatamizadeh2022unetr}. However, variations in imaging acquisition devices, protocols, and patient demographics lead to data distribution discrepancies between training and deployment environments \cite{zhou2022survey}. This mismatch, known as domain shift, severely degrades model performance, compromising the reliability of segmentation algorithms in real-world clinical settings.

Domain generalization (DG) has emerged as a promising approach to mitigating domain shifts in various scenarios \cite{wang2022survey,zhou2022survey}. DG aims to train models on one or multiple related but distinct source domains, enabling them to generalize effectively to unseen target domains. While extensively studied in computer vision, its application in medical image segmentation remains limited due to challenges such as data scarcity and high inter-domain variability. Several DG strategies have been proposed, including domain alignment, meta-learning, and adversarial learning \cite{zhang2023adverin,liu2021feddg,khandelwal2020domain}. Beyond these, large-scale pretraining has proven particularly effective in addressing domain shifts, as models trained on diverse datasets exhibit greater generalization capabilities \cite{kirillov2023segmentanything}. Pretraining on diverse datasets enhances feature robustness and transferability, helping models overcome domain shifts. Actually, the previous research in natural image~\cite{gulrajani2020domainbed} has shown that without the pretrained weights of ImageNet, simple style differences between cartoon and sketch make the classification algorithm fail.  Studies such as CLIP \cite{radford2021clip} demonstrate that zero-shot models trained on diverse datasets are significantly more resilient to distribution shifts than conventional supervised models.  However, large-scale medical image segmentation on MRI remains constrained by the lack of extensive annotated datasets \cite{ma2024medsam,du2023segvol,chen2023samadapter}. The limited availability of annotated medical datasets underscores the need for leveraging unlabeled medical data.

Particularly for MRI images, existing research also primarily focuses on cross-center domain shifts, where models trained on data from one institution are evaluated on data from another \cite{bernard2018acdc,campello2021multi}. While cross-center shifts pose challenges, a largely overlooked yet significant source of domain shift arises from MRI sequence variations, such as differences in echo times and contrast settings under the same modality like T1 or T2, even when imaging the same organ. Unlike cross-center shifts, cross-phase shifts introduce a more complex and clinically relevant challenge, as the vast diversity of imaging sequences makes it impractical to explicitly enumerate or adapt models to every possible variation. Cross-phase settings can reveal totally different domain shift behaviors like cross-center. Despite their critical impact, cross-phase domain shifts remain underexplored due to the lack of comprehensive datasets in the field. 

\noindent \textbf{Contributions}: we summarize key contributions as follows:
\begin{itemize}
\item \textbf{PancreasDG: The first 3D MRI dataset for cross-phase generalization, largest for cross-center shifts.}
We introduce PancreasDG, a novel and challenging 3D MRI segmentation dataset designed to evaluate domain shifts across both data centers and imaging sequences. PancreasDG is the largest dataset for studying cross-center generalization and the first to systematically investigate cross-phase shifts, filling a critical gap in medical image segmentation research.

\item \textbf{Rethink domain shifts in medical segmentation with PancreasDG.}
Our comprehensive experiments on PancreasDG provide new insights into cross-center and cross-phase domain shifts. We show that some observed cross-center differences may stem from case-level variability rather than true distribution shifts. Furthermore, cross-phase shifts exhibit fundamentally different behaviors, challenging assumptions derived from conventional cross-center studies.

\item \textbf{Large-scale semi-supervised pretraining for cross-phase generalization.}
To tackle cross-phase challenges, we propose a semi-supervised learning approach leveraging large-scale unlabeled MRI data. Our model captures diverse anatomical structures, providing a robust initialization for segmentation tasks. Extensive experiments demonstrate that large-scale pretraining significantly improves performance under cross-phase domain shifts, where conventional robustness methods fail. Careful ablation studies further analyze the impact of fine-tuning strategies and prompt variations on final performance.

\end{itemize}
%All data and code are public available for research purposes.

%-------------------------------------------------------------------------
\noindent

\section{Related work}
\label{sec:relevant}

\textbf{Domain generalization:} Domain generalization in medical image segmentation can be broadly categorized into three groups: data manipulation or augmentation, robust representation learning, and other advanced learning strategies like meta-learning or adversarial learning. For example, BigAug~\cite{zhang2020bigaug} shows how stacking multiple data augmentation techniques can significantly enhance segmentation model generalization. Feature manipulation techniques like MixStyle~\cite{zhou2021mixstyle} enable learning more robust features through feature statistic manipulations. ShapeMeta~\cite{liu2020meta} combines meta-learning with shape regularization to ensure the model learns more generalized and robust features. AdverIN, MaxStyle~\cite{zhang2023adverin,chen2020adverbias} employs adversarial attacks to mimic domain bias in medical images, effectively removing domain-specific information. However, the influence of large-scale pertaining to medical segmentation to address the domain shift remains unexplored.

The exploration of domain generalization in MRI datasets remains relatively underdeveloped, primarily due to the scarcity of MRI scans with corresponding annotations. For example, the multi-site dataset for prostate MRI~\cite{liu2020ms} segmentation encompasses T2-weighted MRI data from six different sources, with a focus on intensity variations across centers. Similarly, AMOS\cite{ji2022amos} constructed a dataset aimed at generalizing multi-organ segmentation across CT and T1-weighted MRI scans. The M\&Ms Challenge~\cite{campello2021multi} further introduced a multi-center, multi-disease cardiac segmentation challenge.

\textbf{Large scale medical segmentation:} The Segment Anything Model (SAM)~\cite{kirillov2023segmentanything} has emerged as a first foundational model capable of achieving remarkable segmentation performance across various domains in natural images. Building on this, MedSAM~\cite{ma2024medsam} adapts SAM specifically for medical image segmentation by fine-tuning it on large-scale medical image datasets, covering ten imaging modalities and over 30 cancer types. Further developments include SAM-Med2D~\cite{cheng2023sammed2d}, which extends SAM's utility to two-dimensional medical imaging tasks, providing efficient segmentation in planar medical contexts. Similarly, SAM-Med3D~\cite{wang2024sammed3d} adapts the model to three-dimensional volumetric data with 16 datasets covering diverse medical scenarios, allowing for comprehensive segmentation of complex anatomical structures. SegVol~\cite{du2023segvol} proposes a 3D foundation segmentation model for universal and interactive volumetric medical image segmentation by scaling up training data to 90K unlabeled CT and 6K labeled CT volumes.

\section{Rethink domain shift in medical segmentation via \textit{PancreasDG} dataset.}
\label{sec:method}
\subsection{PancreasDG Dataset}

In this work, we introduce \textit{PancreasDG}, a novel and challenging MRI T1 pancreas segmentation dataset designed to systematically analyze domain shift and evaluate domain generalization algorithms under Institutional Review Board (IRB) approval. The pancreas, a vital abdominal organ with both exocrine and endocrine functions, presents unique segmentation challenges due to its complex anatomical structure and the inherently low signal-to-noise ratio in abdominal MRI scans \cite{zhang2023deep}. Unlike previous datasets, \textit{PancreasDG} uniquely captures two distinct types of domain shifts: (1) Cross-center shifts-Venous-phase T1 MRI scans collected from multiple data centers, enabling the study of inter-center variability (reorganized and expanded from \cite{ZHANG2025large}), and (2) Cross-phase shifts-Newly acquired out-of-phase T1 MRI scans, allowing for the first systematic investigation of segmentation performance across different MRI sequences. 

\begin{figure}[htbp]
    \centering
    \includegraphics[width=\linewidth]{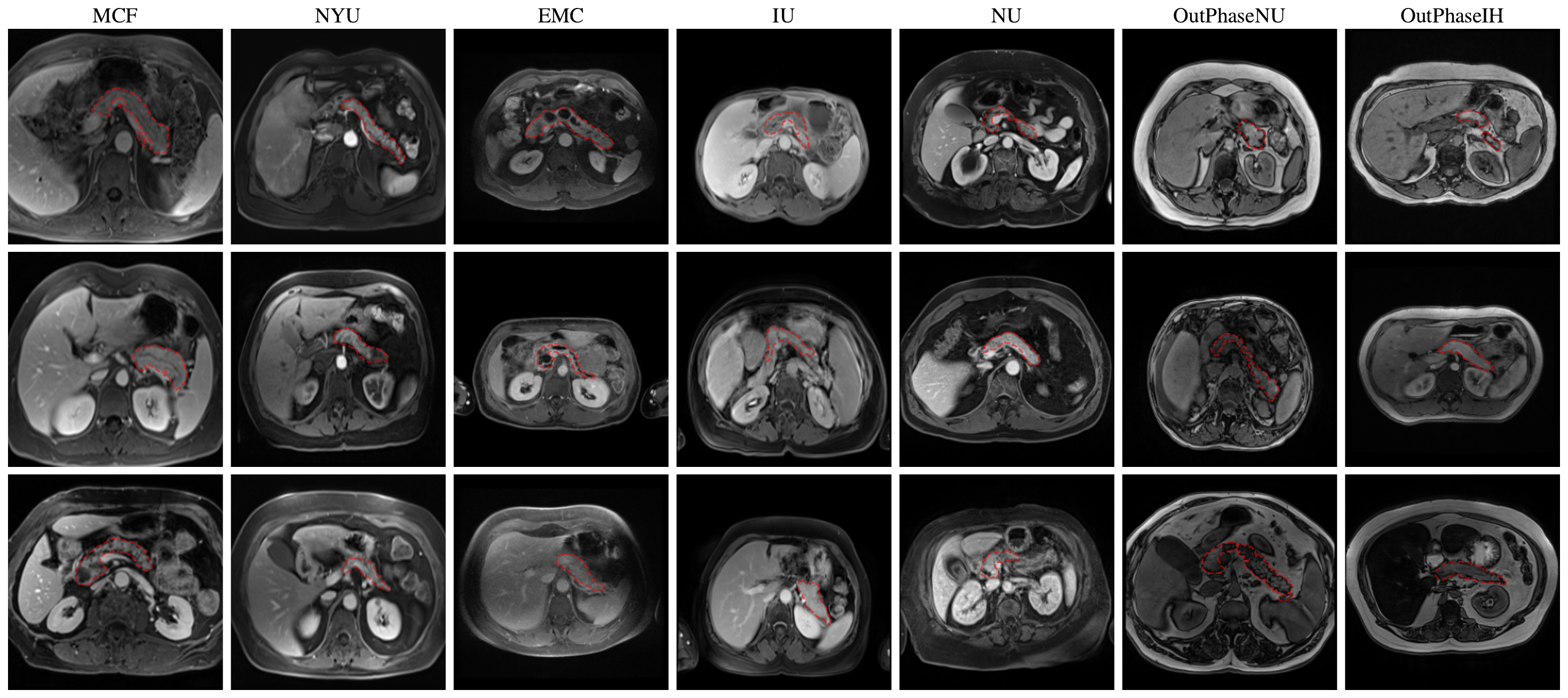}
   \caption{Example MRI scans across different domains for PancreasDG. %There are two kinds of domain shifts that exist in the dataset, one is the distributional shift between different data centers while another one exists for different phases in MRI scans.
    }
    \vspace{-4mm}
    \label{fig:visual_data}
\end{figure}

\textit{PancreasDG} comprises 563 MRI T1 scans collected from six institutions: New York University (NYU) Langone Health, Mayo Clinic Florida (MCF), Northwestern University (NU), Istanbul University Faculty of Medicine (IU), Erasmus Medical Center (EMC), and a private consortium in house data (IH) integrating data from multiple sources, presenting an additional challenge for domain generalization. The overall dataset includes 463 venous-phase scans and 100 out-of-phase scans. A detailed dataset summary is provided in Table~\ref{tab:pancreasdg}, with visual examples in Figure~\ref{fig:visual_data}.

\begin{table}[htbp]
    \caption{We collect the largest ever 3D MRI dataset to investigate domain shifts under cross-center and cross-phase settings. We split this dataset into source domains consisting of venous phase MRI scans, target domain 1 consisting of venous phase scans to investigate cross-center shifts with same sequences, and target domain 2 to investigate the challenging cross-phase shifts.}
    \centering
    \resizebox{0.8\linewidth}{!}{%
    \begin{tabular}{llllll}
        \toprule
        \textbf{Phase} & \textbf{Domain} & \textbf{Center} & \textbf{Devices} & \textbf{Magnet (T)} & \textbf{Samples} \\
        \midrule
        \multirow{5}{*}{Venous Phase} & \multirow{2}{*}{Source} & MCF & GE,Siemens & 1.5, 3 & 151 \\
        \cmidrule(lr){3-6}
        &  & NYU & GE,Siemens & 1.5, 3 & 162 \\
        \cmidrule(lr){2-6} \cmidrule(lr){2-6}
         & \multirow{3}{*}{Target 1} & EMC & GE,Siemens & 1.5 & 50 \\
        \cmidrule(lr){3-6}
        &  & IU & GE, Philips,Siemens & 1.5 & 50 \\
        \cmidrule(lr){3-6}
        &  & NU & Siemens & 1.5, 3 & 50 \\
        \midrule
        \multirow{2}{*}{Out of Phase} & \multirow{2}{*}{Target 2} & NU & Siemens & 1.5, 3 & 50 \\
        \cmidrule(lr){3-6}
        &  & IH & GE, Philips,Siemens & 1.5, 3 & 50 \\
        \bottomrule
    \end{tabular}}
    \label{tab:pancreasdg}
    \vspace{-2mm}
\end{table}

\subsection{Training setups}
To fairly validate the influence of domain shift, we adopt the nnUNet framework which has been validated as the strong segmentation baseline and easy to repeat~\cite{isensee2021nnunet,bassi2024touchstone}. All the training is conducted on one server with 8 80GB A100 GPU. By default, all the following training takes 100 epochs.

Data splits play an important role in domain generalization tasks as shown in natural image domain generalization tasks such as~\cite{gulrajani2020domainbed,zhou2022survey}. However, in the medical segmentation fields, this split might be a slight mess where some works take the target domain data into validation, which could introduce significant bias~\cite{gulrajani2020domainbed}. Thus in this work,  the 313 venous phase cases from source domain data are split into the training, validation, and test according to the ratio of 3:1:1. The remaining 250 cases from various centers including different phases serve as the test set. During the training time, no data from target domains 1 or 2 will participate in any model's validation, making sure that the model is saved according to validation data from the source domain.

\subsection{Domain shift or data lacking?}
%Although previous studies suggest that MRI data acquired from different centers can lead to significant domain shifts, the underlying causes of these shifts and their impact on real-world applications remain unclear due to data limitations for training and testing in prior research~\cite{liu2020ms,campello2021multi}. For example, several centers in multi-site prostate MRI segmentation datasets only contain 12 cases of 3D MRI scans, making it difficult to derive robust conclusions. Leveraging the PancreasDG MRI dataset, the largest 3D MRI dataset for investigating cross-center domain shift, we conduct a comprehensive study on how the size of training and test data influences performance variance across different centers.
The impact of cross-center domain shifts in MRI segmentation remains poorly understood, largely due to data limitations in prior studies~\cite{liu2020ms,campello2021multi}. With most multi-center datasets containing as few as 12 cases per center, conclusions have been tentative at best. Using \textit{PancreasDG}, we systematically investigate how training and test sample sizes affect performance variability across centers, providing more robust insights into the nature of these domain shifts.

\begin{figure}[htbp]
    \centering
    \includegraphics[width=\linewidth]{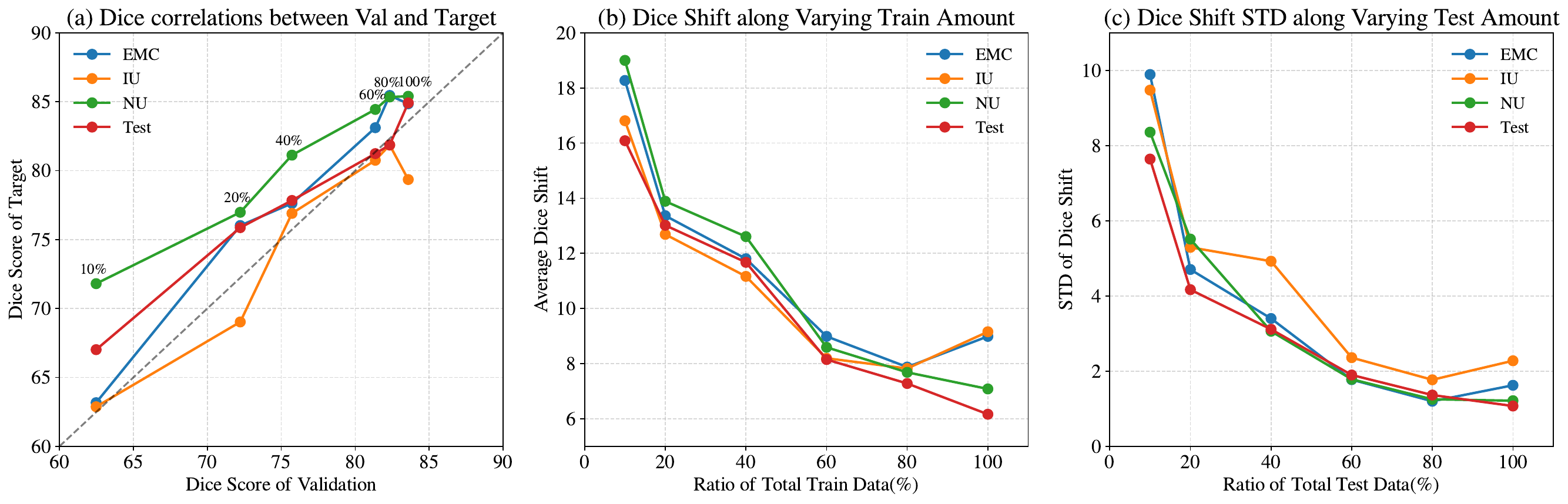}
    \caption{We measure the distribution shift across different centers at the case level between the validation set from the source domain and target data including the test set from the source domain and the test data from the same phase but collected from different data centers along different amount of training and test data. We can observe that along increased number of training data, performance gap between different data centers will be minimized and the increased number of test data can also significantly reduce variance under a cross-center setting.}
    \label{fig:dice_samephase}
    \vspace{-4mm}
\end{figure}

%To systematically assess distribution shifts across centers, we first evaluate segmentation performance variations with varying the amount of training data. Specifically, we compare the performance of target data with the validation set from the source domain at the case level, including the test set from the same source and test data from different centers.  As shown in Figure~\ref{fig:dice_samephase} (a), segmentation performance improves as training data increases, particularly when the dataset is initially small. Interestingly, no substantial difference is observed between the test set from the source domain and data from external centers. Furthermore, Figure~\ref{fig:dice_samephase} (b) illustrates that the performance variance between any target and source validation data decreases significantly with larger training sets. This suggests that the observed Dice score shifts cannot be solely attributed to "domain shifts" but may also stem from case-level variance due to limited training data. This issue is further amplified when test sets are small, as shown in Figure~\ref{fig:dice_samephase}. Large fluctuations in performance emerge which is likely driven by case-specific variance rather than real domain shifts.
To assess cross-center distribution shifts, we analyzed segmentation performance while varying training data size, comparing source domain validation with both source-domain and external center test data. Figure~\ref{fig:dice_samephase}(a) shows that performance improves with more training data across all centers, with external centers exhibiting similar trends to source-domain test sets. Figure~\ref{fig:dice_samephase}(b) reveals that performance variance between target and source data diminishes significantly with larger training sets, suggesting that observed variations may reflect case-level heterogeneity rather than true domain shifts. This effect is amplified with small test sets (Figure~\ref{fig:dice_samephase}(c)), where large performance fluctuations likely stem from case-specific factors rather than systematic domain differences.

%These findings raise critical concerns regarding current domain generalization approaches in medical image segmentation under cross-center settings. Does performance improvement in cross-center experiments truly stem from out-of-distribution generalization, or is it merely a result of better fitting within the source domain? For instance, as demonstrated in Figure~\ref{fig:dice_samephase}, if increasing model parameters improves source domain performance and subsequently enhances performance across centers, should such methods be classified as domain generalization? Similarly, if a data augmentation technique enhances source domain performance while also improving the results of other centers, should it be considered a domain generalization strategy? These questions highlight the need for more rigorous evaluation metrics to accurately quantify cross-center shifts and prevent misleading interpretations in domain generalization research. This is not denying there is a domain shift between different data centers but we need more careful validation when claiming domain generalization in cross-center settings under the same sequences. Overall, these findings emphasize the need for greater caution when defining domain generalization methods in the context of medical image analysis under a cross-center setting. Properly distinguishing between true domain generalization and improvements that primarily benefit the source domain is essential to ensuring meaningful progress in the field.
These findings challenge current approaches to domain generalization in cross-center medical image segmentation. As Figure~\ref{fig:dice_samephase} demonstrates, when improvements in source domain performance directly correlate with enhanced cross-center performance, can this truly be considered domain generalization? The distinction is critical: does a method genuinely address distributional shifts, or simply improve overall performance? While domain shifts between centers certainly exist, our analysis calls for more rigorous evaluation metrics and validation protocols when claiming domain generalization in same-sequence, cross-center settings. Meaningful progress requires clearly distinguishing between methods that enable true out-of-distribution generalization versus those that primarily enhance source domain fitting with corresponding benefits elsewhere.

\subsection{Contradictory results between cross-center and cross-phase shifts along training}
%Building on the previous concerns, traditional segmentation models generally achieve higher final performance with longer training or increased parameter size in the source domain \cite{isensee2021nnunet}. However, such extension training does not necessarily translate to improved performance on out-of-distribution (OOD) data. Normally in natural image, the overfitting in the source domain would even lead to performance drop in the out-of-distribution data~\cite{gulrajani2020domainbed,zhou2022survey}. To investigate this under the medical setting, we conducted a study where we extended model training from 100 epochs to 1000 epochs, saving checkpoints every 100 epochs. We then evaluated the model's performance across different testing centers, considering both same-phase (venous) and cross-phase (out-of-phase) settings. 

While traditional segmentation models achieve better performance with extended training or increased parameter size~\cite{isensee2021nnunet}, these improvements may not transfer to out-of-distribution data. In natural images, source domain overfitting typically reduces performance on OOD data~\cite{gulrajani2020domainbed,zhou2022survey}. To investigate this phenomenon in medical imaging, we extended training from 100 to 1000 epochs, evaluating performance across different test centers for both same-phase (venous) and cross-phase (out-of-phase) settings.

\begin{figure}[htbp]
    \centering
    \includegraphics[width=0.6\linewidth]{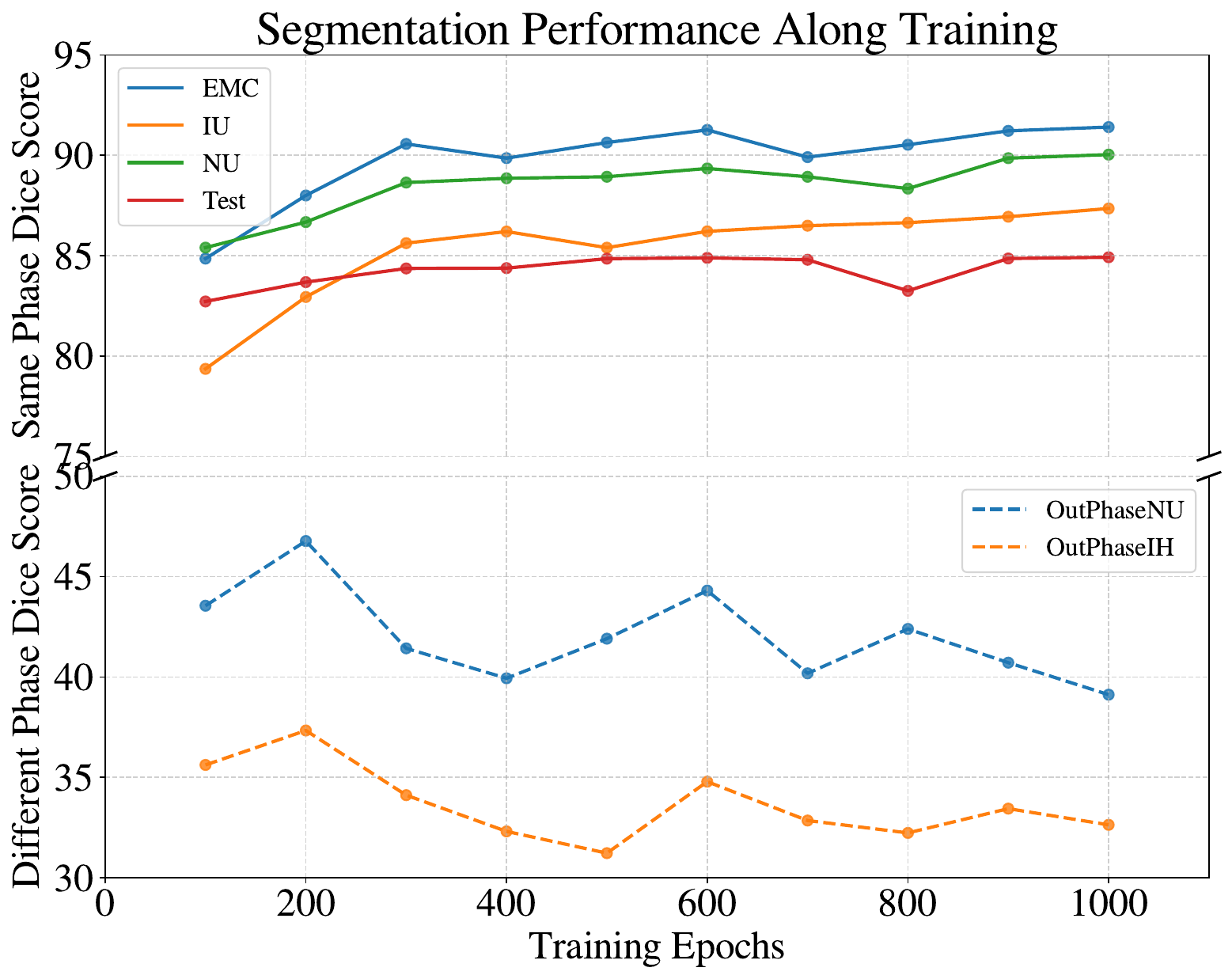}
    \caption{We can observe that along the increasing training epochs, the performance of different data centers with the same imaging phase can share continued performance increasing due to overfitting while the performance with difference phase will drop. }
    \label{fig:seg_along_train}
\end{figure}
%As shown in Figure~\ref{fig:seg_along_train}, regardless of training duration, models consistently improve performance within the same phase, and data from different centers can even take more advantage than the test data from the source domain. However, in the cross-phase setting, performance gains remain minimal, and actually, extended training even results in performance degradation. This observation highlights a fundamental difference between cross-center and cross-phase domain shifts. While models can adapt to cross-center variations with sufficient training, cross-phase domain shifts present a more complex challenge, where simple accuracy improvements in the source domain do not guarantee better generalization to the target domain. Further experiments Figure~\ref{fig:seg_along_block} also shows that increasing the model parameters can lead to performance advancement same sequence data regardless of center, but the performance can even degrade under cross-phase data. These findings align with previous study in natural image studies \cite{gulrajani2020domainbed,zhou2022survey}, where performance improvements in the training domain do not necessarily lead to improved generalization.

Figure~\ref{fig:seg_along_train} reveals a striking contrast: with extended training, performance consistently improves for same-phase data across all centers, yet deteriorates for cross-phase data. This highlights a fundamental difference between cross-center and cross-phase domain shifts—while models adapt to center variations with sufficient training, cross-phase shifts present a more complex challenge where source domain improvements fail to generalize. Similarly, Figure~\ref{fig:seg_along_block} shows that increasing model capacity benefits same-sequence performance regardless of center, but degrades cross-phase performance. These findings align with natural image domain generalization studies \cite{gulrajani2020domainbed,zhou2022survey}, where training domain improvements often fail to translate to out-of-distribution generalization.

Moreover, we observe that the Dice score variance in cross-phase settings remains significant throughout training, emphasizing the importance of carefully isolating target domain data during validation and model selection. These findings suggest that cross-phase domain shifts introduce a fundamentally different generalization challenge compared to traditional simple cross-center shifts, warranting more robust evaluation strategies in future research.

\section{Semi-supervised Learning Approach}
\begin{figure}[htbp]
    \centering
    \includegraphics[width=\linewidth]{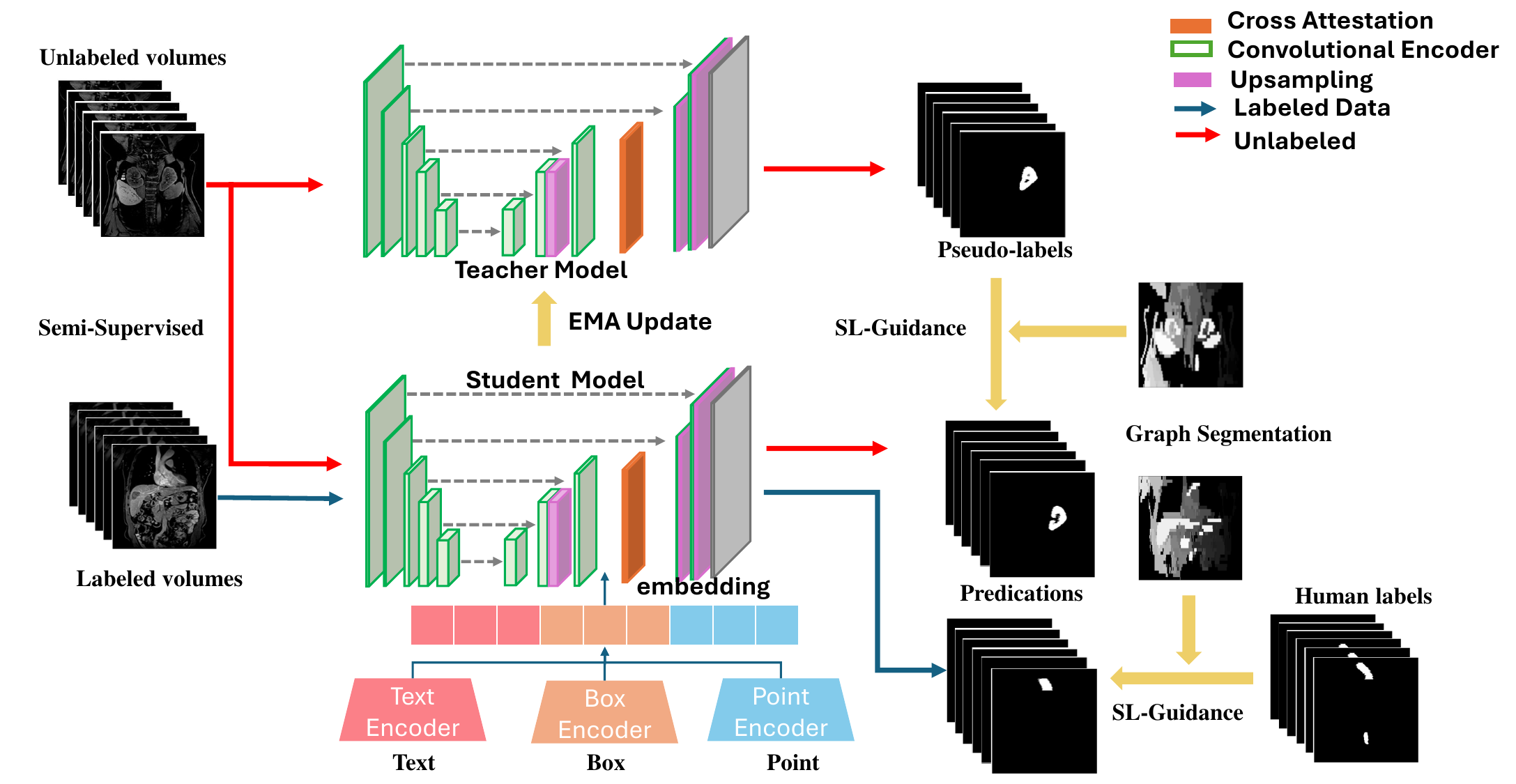}
    \caption{We adopt the EMA Network structure for the semi-supervised pretraining on the unlabeled cases. To avoid the segmentation collapse during the training, we add the graph segmentation result as the regularization such that we can further extend the model's performance on the general segmentation tasks.}
    \label{fig:sam_semi}
    \vspace{-4mm}
\end{figure}

\subsection{Large scale pretraining on nnUNet}
%Previous research has highlighted the dominance of the nnUNet training framework~\cite{isensee2021nnunet}, which consistently outperforms other architectures, such as MONAI~\cite{cardoso2022monai}, in medical image segmentation tasks~\cite{bassi2024touchstone}. Despite its leading performance, large-scale pretraining on the nnUNet architecture remains underexplored, particularly for 3D MRI images. This limitation is primarily due to the scarcity of annotated datasets, which poses a significant barrier to pretraining on a large scale. An example of progress in this area is STU-Net\cite{huang2023stunet}, which performs scalable and transferable medical image segmentation using 1,204 CT images with 104 anatomical structures from the TotalSegmentator dataset\cite{wasserthal2023totalsegmentator}, building on the nnUNet framework. However, these methods are constrained by their reliance on fixed classification categories, limiting their ability to incorporate diverse segmentation targets across different modalities.

The nnUNet framework~\cite{isensee2021nnunet} dominates medical image segmentation benchmarks~\cite{cardoso2022monai}, yet large-scale pretraining on this architecture remains underexplored for 3D MRI due to annotated data scarcity. While STU-Net~\cite{huang2023stunet} has demonstrated transferable segmentation using 1,204 CT scans from TotalSegmentator~\cite{wasserthal2023totalsegmentator}, such approaches remain constrained by fixed classification categories, limiting their ability to integrate diverse segmentation targets across modalities—a critical requirement for addressing cross-sequence domain shifts.

%This fixed-category approach restricts the flexibility needed for integrating varied datasets, making it challenging to leverage the full diversity of medical imaging data. Inspired by recent advancements in general-purpose segmentation models like the SAM~\cite{kirillov2023segmentanything} and SegVol~\cite{du2023segvol}, a new paradigm for medical image segmentation has emerged. By introducing embeddings derived from text prompts, bounding box coordinates, or point annotations as input and generating segmentation masks as output, a unified framework can be established. This approach enables the seamless integration of datasets from different imaging modalities, device configurations, and segmentation targets. Such a framework facilitates the training of robust, generalizable models capable of addressing the diverse challenges inherent in medical image segmentation, advancing the field toward universal and adaptable solutions.
Fixed-category approaches limit cross-dataset integration in medical imaging. Inspired by SAM~\cite{kirillov2023segmentanything} and SegVol~\cite{du2023segvol}, we adopt a more flexible paradigm that accepts various inputs (text prompts, bounding boxes, or point annotations) to generate segmentation masks. This framework enables seamless integration of diverse imaging modalities and segmentation targets, facilitating more robust and generalizable models that can better address cross-sequence domain shifts.

In this study, we leverage 25 open-source CT segmentation datasets referenced in~\cite{du2023segvol}. %, including CHAOS, HaN-Seg, AMOS22, AbdomenCT-1k, KiTS23, KiPA22, KiTS19, BTCV, Pancreas-CT, 3D-IRCADB, FLARE22, TotalSegmentator, CT-ORG, VerSe19, VerSe20, SLIVER07, QUBIQ, six MSD datasets, LUNA16, and WORD. 
In addition, we collect 10 MRI datasets, including BRATS~\cite{menze2014brats}, TotalSegmentatorMRI~\cite{d2024totalsegmentatormri}, AMOS22-MRI~\cite{ji2022amos}, CirrMRI600~\cite{jha2024cirrmri600}, Duke Liver~\cite{macdonald2023dukemri}, and ACDC Cardiac MRI~\cite{bernard2018acdc}, to establish a comprehensive dataset pool spanning both CT and MRI imaging modalities. As we can observe, the span of segmentation category and data amount in MRI are much more limited compared to CT. We apply over-sampling to MRI segmentation data to ensure a balanced training process, maintaining a 1:1 training ratio between MRI and CT datasets.

To address the scarcity of annotated MRI data, we leverage 6.5K unlabeled 3D MRI scans from diverse centers and sequences using a semi-supervised approach (Figure~\ref{fig:sam_semi}). Our method employs an Exponential Moving Average (EMA) teacher-student framework, where the teacher model generates pseudo-labels from randomly sampled foreground points in unlabeled scans. This approach encourages flat optimization minima, improving generalization across domain shifts. To prevent convergence to trivial solutions, we incorporate unsupervised graph-based segmentation via the Felzenszwalb-Huttenlocher algorithm [11,14] as a regularization signal. This combined framework exposes the model to diverse imaging characteristics while maintaining anatomical consistency, enhancing robustness for cross-sequence generalization. Pretraining required 10 epochs with a learning rate of $10^{-5}$, completing in approximately one week.

\subsection{Finetuning on the PancreasDG dataset}
Given the limitations of publicly available annotated MRI datasets, our objective in this work is not to develop a universal segmentation model capable of handling all scenarios, as such an approach is neither realistic nor practical. Prior research has also demonstrated that generalized segmentation models often fail to match the performance of task-specific models in the medical domain~\cite{bassi2024touchstone}. Instead, we leverage large-scale semi-supervised training as a robust foundation for downstream segmentation tasks. To this end, we fine-tune our model using pretrained weights from the initial training steps within the nnUNet framework, ensuring adaptability to specific segmentation applications. Finetuning on the PancreasDG dataset takes 100 steps with the standard nnUNet training with a small learning rate of $10^{-4}$.

\subsection{Result comparison}
\label{sec:results}

\begin{table}[htbp]
\vspace{-4mm}
\caption{Segmentation performance on the out-of-phase NU dataset is shown, following training exclusively on venous-phase data. Results illustrate the limitations of current state-of-the-art methods in handling severe cross-sequence domain shifts, none of which achieve satisfactory results. Notably, the proposed model, equipped with large-scale semi-supervised pretraining, demonstrates substantial performance gains.}
    \centering
    \resizebox{0.96\linewidth}{!}{
    \begin{tabular}{lcccccc}
    \toprule
     & Dice & Jaccard & Precision & Recall & HD95 (mm) & ASSD (mm) \\
     \midrule
    Baseline & 43.55 ± 26.33 & 31.39 ± 21.31 & 55.79 ± 32.77 & 37.54 ± 24.11 & 39.88 ± 26.43 & 13.23 ± 10.73 \\
    \midrule
    \multicolumn{7}{c}{\textit{Domain Generalization Methods}}\\
    \midrule
    EQRM \cite{eastwood2022probable} & 44.68 ± 25.73 & 32.18 ± 20.78 & 63.42 ± 31.61 & 36.78 ± 22.72 & 36.34 ± 25.45 & 11.49 ± 9.97 \\
    GroupDRO \cite{sagawa2019distributionally} & 43.46 ± 26.50 & 31.29 ± 21.04 & 61.98 ± 34.03 & 35.67 ± 23.20 & 41.27 ± 33.24 & 14.41 ± 16.46 \\
    IBerm \cite{ahuja2021invariance} & 43.55 ± 26.33 & 31.36 ± 21.16 & 61.32 ± 32.12 & 35.87 ± 23.56 & 38.47 ± 30.84 & 12.85 ± 12.57 \\
    BigAug~\cite{zhang2020bigaug} & 40.43 ± 26.09 & 28.65 ± 20.49 & 57.21 ± 34.60 & 32.52 ± 22.35 & 42.37 ± 26.93 & 15.71 ± 15.77 \\
    RandConv~\cite{xu2021randcov} & 43.59 ± 26.19 & 31.33 ± 20.89 & 54.07 ± 32.29 & 37.68 ± 23.45 & 39.43 ± 26.50 & 13.04 ± 11.21 \\
    SD \cite{pezeshki2021gradient} & 40.59 ± 26.45 & 28.97 ± 21.33 & 56.75 ± 34.03 & 33.72 ± 23.53 & 42.95 ± 28.96 & 14.46 ± 12.46 \\
    VREX \cite{krueger2021out} & 46.78 ± 24.49 & 33.69 ± 19.94 & 65.22 ± 28.90 & 39.07 ± 22.33 & 36.40 ± 26.70 & 11.26 ± 10.15 \\
    MixStyle \cite{zhou2021mixstyle} & 49.71 ± 20.35 & 35.34 ± 16.91 & 57.31 ± 24.62 & 45.89 ± 19.67 & 43.30 ± 29.89 & 10.78 ± 7.63 \\
    \midrule
    \multicolumn{7}{c}{\textit{Large Segmentation Model}}\\
    \midrule
    SAM \cite{kirillov2023segment} & 36.58 ± 10.52 & 22.91 ± 8.22 & 23.31 ± 8.43 & 95.20 ± 7.96 & 35.65 ± 9.58 & 13.14 ± 4.43 \\
    SAM-Med2D \cite{cheng2023sam} & 9.72 ± 4.17 & 5.16 ± 2.34 & 16.78 ± 11.92 & 8.55 ± 3.95 & 63.08 ± 36.75 & 22.35 ± 15.73 \\
    SegVol \cite{du2025segvol} & 32.80 ± 15.17 & 20.66 ± 11.59 & 54.10 ± 23.88 & 28.54 ± 16.80 & 32.71 ± 14.73 & 10.78 ± 5.70 \\
    \midrule
    \multicolumn{7}{c}{\textit{Pretrained Medical Segmentation Model}}\\
    \midrule
    STUNet-Base \cite{huang2023stu} & 37.68 ± 26.35 & 26.53 ± 20.63 & 56.23 ± 35.12 & 30.05 ± 22.50 & 40.04 ± 25.56 & 14.36 ± 12.49 \\
    STUNet-Small \cite{huang2023stu} & 32.55 ± 27.18 & 22.81 ± 20.97 & 56.02 ± 37.57 & 25.22 ± 23.00 & 49.82 ± 30.46 & 20.24 ± 17.43 \\
    STUNet-Large \cite{huang2023stu} & 36.29 ± 26.90 & 25.61 ± 21.16 & 56.38 ± 37.01 & 28.58 ± 22.91 & 43.92 ± 27.81 & 16.31 ± 13.49 \\
    \midrule
    Ours & \textbf{70.39 ± 7.82} & \textbf{54.86 ± 9.07} & \textbf{83.25 ± 9.44} & \textbf{61.92 ± 9.47} & \textbf{13.04 ± 9.28} & \textbf{3.40 ± 1.92} \\
    \bottomrule
    \end{tabular}}
    
    \label{tab:outphasenu_performance}
\vspace{-8mm}
\end{table}

\begin{table}[htbp]
\vspace{-4mm}
\caption{The segmentation performance in the out of phase IH dataset while the training is only conducted in the venous phase. We can observe a similar trend that our pretrained model significantly outperforms other domain generalization methods. }
    \centering
    \resizebox{0.75\linewidth}{!}{
    \begin{tabular}{lcccccc}
    \toprule
     & Dice & Jaccard & Precision & Recall & HD95 & ASSD\\
     \midrule
    Baseline & 35.62 & 26.75 & 41.12 & 33.19 & 46.10 & 17.75 \\
    \midrule
    \multicolumn{7}{c}{\textit{Domain Generalization Methods}}\\
    \midrule
    EQRM \cite{eastwood2022probable} & 36.11 & 26.79 & 47.17 & 31.87 & 36.80 & 14.31 \\
    GroupDRO \cite{sagawa2019distributionally} & 35.96 & 26.71 & 47.48 & 31.17 & 35.42 & 13.94 \\
    IBerm \cite{ahuja2021invariance} & 35.22 & 26.09 & 43.84 & 31.87 & 43.68 & 16.36 \\
    BigAug~\cite{zhang2020bigaug} & 37.15 & 27.26 & 44.22 & 35.59 & 44.09 & 16.85 \\
    RandConv~\cite{xu2021randcov} & 36.30 & 26.70 & 40.76 & 35.01 & 42.75 & 16.88 \\
    SD \cite{pezeshki2021gradient} & 32.03 & 23.93 & 41.97 & 27.39 & 40.57 & 16.38 \\
    VREX \cite{krueger2021out} & 38.10 & 28.08 & 46.65 & 34.12 & 40.10 & 14.57 \\
    MixStyle \cite{zhou2021mixstyle} & 43.25 & 31.61 & 45.50 & 44.02 & 45.82 & 13.69 \\
    \midrule
    \multicolumn{7}{c}{\textit{Large Segmentation Model}}\\
    \midrule
    SAM \cite{kirillov2023segment} & 35.35 & 22.44 & 22.52 & 99.23 & 31.95 & 12.76 \\
    SAM-Med2D \cite{cheng2023sam} & 12.90 & 7.29 & 19.66 & 16.80 & 78.60 & 33.79 \\
    SegVol \cite{du2025segvol} & 45.81 & 30.74 & 54.03 & 46.90 & 18.75 & 6.29 \\
    \midrule
    \multicolumn{7}{c}{\textit{Pretrained Medical Segmentation Model}}\\
    \midrule
    STUNet-Base \cite{huang2023stu} & 30.69 & 22.07 & 42.27 & 26.48 & 43.37 & 17.41 \\
    STUNet-Small \cite{huang2023stu} & 28.20 & 20.35 & 46.79 & 23.24 & 50.45 & 20.78 \\
    STUNet-Large \cite{huang2023stu} & 26.01 & 18.67 & 39.40 & 21.15 & 48.44 & 20.45 \\
    \midrule
    Ours & \textbf{66.61} & \textbf{52.46} & \textbf{73.26} & \textbf{63.33} & \textbf{15.68} & \textbf{4.54} \\
    \bottomrule
    \end{tabular}}
    
    \vspace{-4mm}
    \label{tab:outphasecpdpc_performance}
\end{table}

%The results in Table~\ref{tab:outphasenu_performance} show the performance of different segmentation methods including various domain generalization methods, the large segmentation models, and the model with pretrained weights on the NU data center. We can clearly observe that some domain generalization methods can achieve significant performance gains compared with the strong nnUNet segmentation baseline. However, those performances remain limited to achieve one satisfactory segmentation performance, and the best performance domain generalization method MixStyle can achieve the performance from 43.55 to 49.71. Compared with these methods, our proposed method can lead to one fundamental change to 70.39 in terms of Dice coefficient. The following Figure~\ref{fig:semiseg_visual} provides one intuitive comparison between segmentation performance with or without large-scale semisupervised pre-training. Overall, our results underscore the potential of large-scale pretraining in bridging the gap between different imaging phases, thereby improving diagnostic accuracy and robustness across varied medical scenarios. This approach not only advances the field of medical image segmentation but also sets a precedent for future research into semi-supervised learning models in healthcare applications. 
Table~\ref{tab:outphasenu_performance} compares segmentation performance on the out-of-phase NU dataset. While existing DG methods and large segmentation models provide some gains over the nnU-Net baseline (Dice 43.55\%), their performance remains limited (e.g., MixStyle at 49.71\% Dice). In contrast, the proposed semi-supervised pretraining achieves a markedly superior Dice score of 70.39\%. This significant improvement, also evident in the qualitative results in Figure~\ref{fig:semiseg_visual}, underscores the power of large-scale pretraining to effectively bridge challenging cross-sequence domain gaps. The proposed approach enhances segmentation robustness for varied medical scenarios and demonstrates a promising direction for semi-supervised learning in healthcare.

% \begin{figure}[htbp]
%     \centering
%     \includegraphics[width=0.6\linewidth]{figs/OutPhaseNU_Segmentation_Visualization.pdf}
%     \caption{The segmentation performance comparison on out-of-phase data between our proposed method and other domain generalization methods.}
%     \vspace{-4mm}
%     \label{fig:semiseg_visual}
% \end{figure}

\section{Ablation studies}
\textbf{Influence of fine-tuning blocks:}
% \begin{figure}[htbp]
%     \centering
%     \includegraphics[width=0.6\linewidth]{figs/domain_performance_along_blocks.pdf}
%     \caption{We can observe that along the number of fine-tuning blocks and increasing parameter space, the performance of different data centers with the same phase will increase while the performance of the difference phase will drop. }
%     \label{fig:seg_along_block}
%     \vspace{-4mm}
% \end{figure}
We investigate the influence of fine-tuning different numbers of blocks in the network, as this affects the parameter space available for adaptation. As shown in Figure 7, with increasing numbers of fine-tuning blocks and growing parameter space, the performance on different data centers with the same phase improves, while performance on different phases decreases. This highlights another difference between cross-center and cross-phase domain shifts: there exists a trade-off between performance on same-phase and different-phase data. According to the results, fine-tuning the last two blocks achieves the best balance between performance on same-phase and different-phase data, suggesting an optimal strategy for model adaptation.

%The influence of different fine-tuning blocks also exists under investigation, given that it increases the parameter size for fine-tuning. As shown in the previous natural image classification tasks, researchers will try to fix the image encoder and only fine-tune the last classification head in the domain generalization problem~\cite{gulrajani2020domainbed}, and fine-tuning the whole network will lead to a significant performance drop in the out-of-distribution domain. However, the influence of different fine-tuning blocks for segmentation networks like UNet-style structures remains unexplored. Here, we compare the model's performance with increasing fine-tuning blocks from the segmentation head in the network decoder. Here, we can clearly observe that with increasing the number of fine-tuning blocks and increasing parameter space, the performance of different data centers with the same phase will increase while the performance of the difference phase will drop, which again raise the attention for different performance between cross-center and cross-phase shift. There exists one balance between the performance from the same phase and different phases regardless of data centers. According to the results, fine-tuning the last two blocks achieve the best performance in this task.

\textbf{Influence of different prompts:} We also investigate the impact of different prompts (points, boxes) on segmentation performance. While one might suspect that additional prompts help the model achieve more reliable results against domain shifts, our results in Table~\ref{tab:outphasenu_performance} show that even without additional prompts, our fine-tuned model based on the pretrained network achieves superior performance compared to other domain generalization methods.  Adding points or boxes as prompts further improves performance, with bounding box prompts providing the most significant benefit. Interestingly, combining both point and box prompts does not lead to further improvements, suggesting that these prompts may provide redundant information for the model.

%One suspect for the performance gain might be coming from introducing the points or box, which help the model to achieve more reliable results against the domain shift. This is partly true while we can also observe that when increasing the prompts like points and boxes can further increase the segmentation performance, the fine-tuned model without additional prompts can still achieve one superior segmentation performance comparing with other domain generalization methods as listed in Table~\ref{tab:outphasecpdpc_performance} and Table~\ref{tab:outphasenu_performance}.
\begin{table}[htbp]
\caption{Influence of different prompts on segmentation performance.}
    \centering
    \resizebox{0.6\linewidth}{!}{
    \begin{tabular}{ccccc}
    \toprule
     & \multicolumn{2}{c}{NU} & \multicolumn{2}{c}{IH} \\
     \cmidrule(lr){2-5}
     & Dice & HD95 & Dice & HD95\\
    \midrule
    Baseline & 43.55 & 39.88 & 35.62 & 46.10 \\
    \midrule
    No add. prompts & 65.70 & 28.28 & 60.95 & 23.72 \\
    \midrule
    + Points & 67.78  & 15.89 & 61.73  & 19.30  \\
    + Box & 70.39 & 13.04  & 66.61 & 15.68  \\
    + Point + Box & 68.89  & 14.75  & 65.10  & 16.82  \\
    \bottomrule
    \end{tabular}}
    
    \vspace{-4mm}
    \label{tab:embedding_comparision}
\end{table}

\section{Discussion and Conclusion}
\label{sec:conlusion}
%Overall, in this work, we propose one largest but challenging PancreasDG MRI benchmarks to thoroughly investigate domain shifts under real clinical setups. Based on the proposed dataset, we observe that the cross-phase setting provides a different domain shift compared with the traditional cross-center setting. Many observations like the trade-off between source and target domain data are also observed under cross-phase setting, which aligns with the domain generalization studies on the natural image analysis. We further show that through we can significantly improve the model's performance with our proposed large-scale semi-supervised pertaining.
We present \textit{PancreasDG}, a new public benchmark resource comprising 563 multi-center, multi-sequence 3D MRI scans, specifically curated to investigate domain generalization for pancreas segmentation. \textit{PancreasDG} uniquely enables the systematic study of both cross-center and, for the first time on this scale, critical cross-sequence (e.g., venous vs. out-of-phase T1) domain shifts. Our analysis reveals that cross-sequence variations pose distinct and often more severe challenges than cross-center shifts with consistent sequences, urging the community to consider these complex scenarios in developing robust clinical AI. Furthermore, we highlight the importance of distinguishing true domain shifts from performance variations due to limited data sampling. Accompanying this dataset, we introduced a powerful semi-supervised pretraining algorithm that effectively learns sequence-invariant anatomical representations from large unlabeled MRI collections. Our method demonstrates remarkable improvements in cross-sequence pancreas segmentation, outperforming existing domain generalization techniques and large-scale segmentation models when faced with these challenging shifts. We believe \textit{PancreasDG} will serve as a crucial catalyst for future research in domain generalization. The dataset, along with our proposed pretraining approach, provides the tools and insights necessary to develop and rigorously evaluate the next generation of robust medical image segmentation models, ultimately paving the way for more reliable clinical translation.

While \textit{PancreasDG} represents a significant advancement in medical image segmentation datasets, it has several limitations that should be acknowledged: \textbf{(A) Limited Sequence Diversity}: although we include both venous phase and out-of-phase sequences, the diversity of MRI sequences in clinical practice is much greater. Future extensions should incorporate additional sequences such as T2-weighted, diffusion-weighted, and arterial phase images. \textbf{(B) Annotation Variability}: Despite our quality control procedures, inter-observer variability in pancreas segmentation remains a challenge. This variability may affect the ground truth quality, particularly for cases with poor image contrast or complex anatomy. \textbf{(C) Single Organ Focus}: While pancreas segmentation presents a challenging test case, domain generalization strategies might perform differently for other anatomical structures. Future work should explore multi-organ segmentation to assess the generalizability of our findings.

\newpage

\small
\bibliographystyle{ieeenat_fullname}
\bibliography{main}

\appendix

\section{Technical Appendices and Supplementary Material}
\begin{figure}[htbp]
    \centering
    \includegraphics[width=0.6\linewidth]{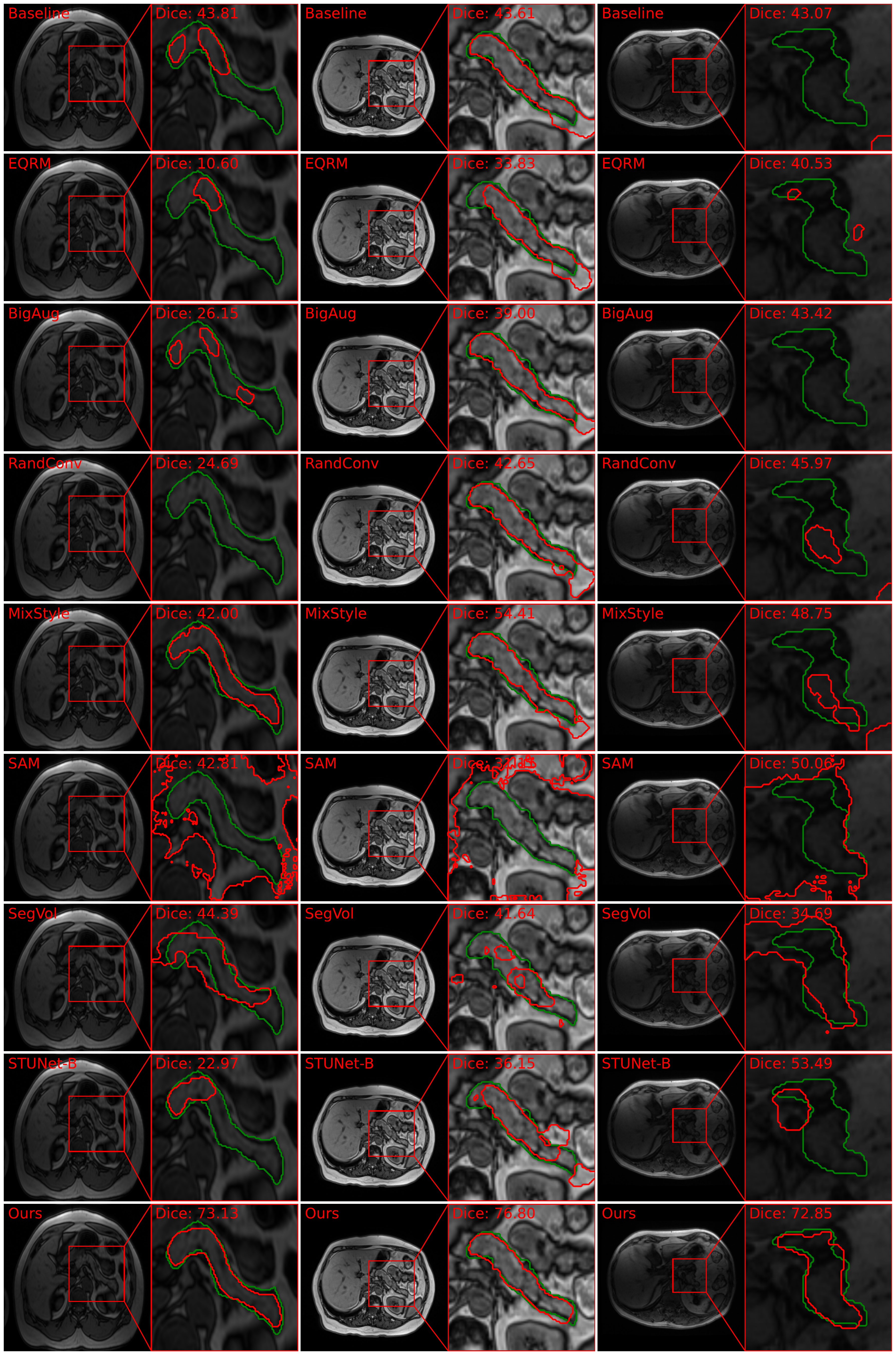}
    \caption{The segmentation performance comparison on out-of-phase data between our proposed method and other domain generalization methods.}
    \vspace{-4mm}
    \label{fig:semiseg_visual}
\end{figure}

\begin{figure}[htbp]
    \centering
    \includegraphics[width=0.6\linewidth]{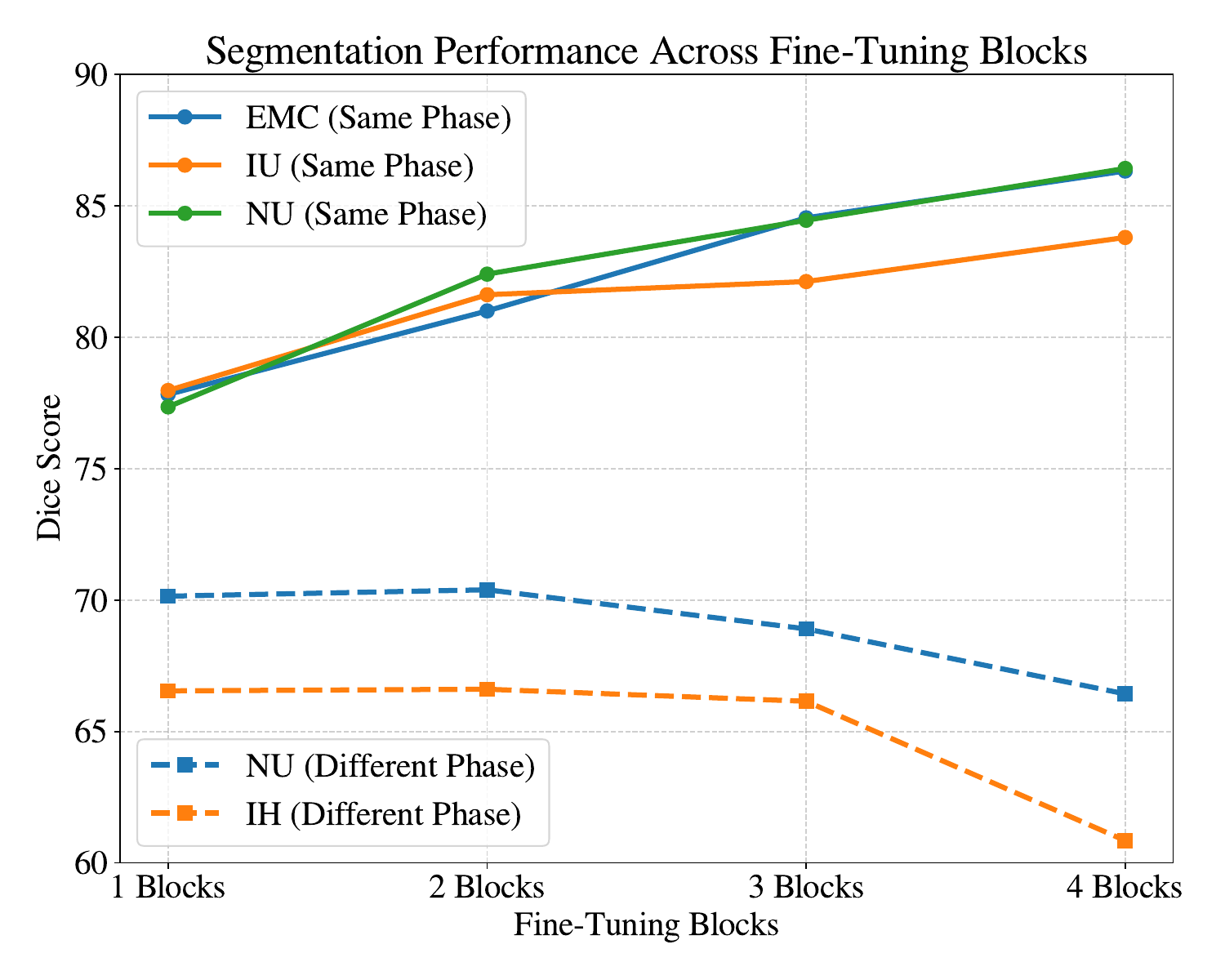}
    \caption{We can observe that along the number of fine-tuning blocks and increasing parameter space, the performance of different data centers with the same phase will increase while the performance of the difference phase will drop. }
    \label{fig:seg_along_block}
    \vspace{-4mm}
\end{figure}

\end{document}